\documentclass{ajour}
\usepackage{amsmath, epsfig}
\input{amssym.def}
\newsymbol\lesssim 132E
\newsymbol\gtrsim 1326
\newcommand{\dphi}{\frac{d\phi}{d\ln r}}
\newcommand{\ydef}{\frac{p}{p+\rho}}

\begin{document}

\authorrunninghead{U. S. Nilsson and C. Uggla}
\titlerunninghead{General Relativistic Stars}

\title{General Relativistic Stars : Polytropic Equations of State}

\author{Ulf S. Nilsson}
\affil{Department of Applied Mathematics\\University of
  Waterloo\\Waterloo, Ontario\\Canada, N2L 3G1\\and\\
Department of Physics\\
Stockholm University\\
Box 6730\\
S-113 95 Stockholm\\
Sweden}
\email{unilsson@math.uwaterloo.ca}

\and

\author{Claes Uggla}
\affil{Department of Physics\\University of Karlstad\\S-651 88
Karlstad\\Sweden}
\email{uggla@physto.se}

\abstract{In this paper, the gravitational field equations for static 
  spherically 
  symmetric perfect fluid models with a polytropic equation of state,
  $p=k\rho^{1+1/n}$, are recast into two complementary 3-dimensional 
  {\it regular} systems of ordinary differential equations on compact state 
  spaces. The systems are analyzed numerically and qualitatively,
  using the theory of dynamical systems. Certain key solutions are shown to 
  form building blocks which, to a large extent, determine the remaining
  solution structure. In one formulation, there exists a monotone function 
  that forces the general relativistic solutions towards a part of the 
  boundary of the state space that corresponds to the low pressure limit. The
  solutions on this boundary describe Newtonian models and thus the 
  relationship to the Newtonian solution space is clearly displayed. 
  It is numerically demonstrated that general relativistic models have
  finite radii when the  
  polytropic index $n$ satisfies $0\leq n \lesssim 3.339$ and infinite radii 
  when $n\geq 5$. When $3.339\lesssim n<5$, there exists a 1-parameter
  set of models with finite radii and a finite number, depending on
  $n$, with infinite radii.} 








\begin{article}

\section{Introduction}
This paper is the second in a series dealing with general relativistic
star models. Here we consider static spherically symmetric models.
The line element for these models can be written as 
\begin{equation}
\label{eq:ds2}
 ds^2 = -{\rm e}^{2\phi(\lambda)}dt^2 +
 r(\lambda)^2\left[\tilde{N}(\lambda)^2d\lambda^2 + d\Omega^2\right] \ ,
\end{equation}
with
\begin{equation}
d\Omega^2 = d\theta^2 + \sin^2\theta d\varphi^2\ ,
\end{equation}
where $\phi(\lambda)$ is the gravitational potential, $r(\lambda)$ is the usual
Schwarzschild radial parameter, and $\tilde{N}(\lambda)$
a dimensionless (under scale-transformations) freely specifiable function.
The choice $\tilde{N}=1$ corresponds to isotropic coordinates and the function 
$\tilde{N}$ can hence be viewed as a relative gauge function with respect to 
the isotropic gauge. The coordinate $\lambda$ is a spatial radial 
variable, defined by the choice of $\tilde{N}$.

The matter content of the star is assumed to be a
perfect fluid, described by the energy-momentum tensor 
\begin{equation}
 \label{eq:emom}
 T_{ab} = \rho u_a u_b + p\left(g_{ab} + u_a u_b \right)\ ,
\end{equation}
where $\rho$ is the energy density, $p$ the pressure, and
$u^a$ the 4-velocity of the fluid.
In this paper we focus on polytropic equations of state
\begin{equation}
  \label{eq:polyeqstate}
  p = k\rho^\Gamma\ ,
\end{equation}
where the constant $\Gamma$ is related to the polytropic 
index $n$ according to $\Gamma = 1 + 1/n$. It will be assumed that
$n\geq0$, where the case $n=0$ corresponds to an incompressible
fluid. In the limit $n\rightarrow\infty$, the polytropic equation of state 
(\ref{eq:polyeqstate}) becomes linear and scale-invariant (it thus corresponds 
to the special case $\rho_0=0$ in the linear equation of state,
$\rho=\rho_0 + (\eta-1)p$, treated in \cite{art:grstar_leq}).

Newtonian polytropic models have been studied for over a hundred years.
Early results have been extensively described by Chandrasekhar
\cite{book:Chandra1939}, however, it is still an active area of research as 
exemplified by the fairly recent papers by Kimura \cite{art:Kimura1981} and 
Horedt \cite{art:Horedt1987}. This indicates that the
Newtonian case is quite non-trivial, and the relativistic case turns
out to be even more 
complicated, as we will demonstrate in this paper.
The history of the relativistic case is also quite long.
For example, Tooper \cite{art:Tooper1964} studied these models 35 years ago, 
but they have also been studied recently by, for example, deFelice {\em et al} 
\cite{art:deFeliceetal1995}. Often polytropes describe low or high pressure
regimes of more realistic equations of state for white dwarfs and 
neutron stars. These regimes decide many of the models physical features,
which therefore can be understood if the corresponding polytropic models are 
understood. One can also construct physically realistic composite equations 
of state by letting certain pressure
regimes be described by polytropic equations of state.
There are thus ample reasons to study models with polytropic equations of 
state, and it is not surprising that these models have attracted attention 
for a long time.

Substituting (\ref{eq:ds2}) into the equations of motion of the fluid, 
$\nabla_a T^{ab}=0$, relates the gravitational potential $\phi$ in
(\ref{eq:ds2}) to the matter content of the spacetime according to
\begin{equation}
 \label{eq:dphidp} 
 \frac{d\phi}{dp} = -\frac{1}{\rho+p}\ .
\end{equation}

To facilitate the study of the gravitational field equations, we
introduce various sets of bounded dimensionless variables (under 
scale-transformations), which regularize the gravitational field equations. 
Recasting the 
equations into regularized form on a compact state space allows us to
numerically and qualitatively explore the solution space, using  
methods from dynamical systems theory. It is worth noticing that one can 
modify the approaches we introduce and apply them to large classes of
equations of state. The dynamical systems are 3-dimensional in
all cases. This is a great advantage since one therefore can {\em visualize} 
the compact state space and obtain a picture of the structure of the 
{\em entire} solution space for large classes of equations of state. 

In the interior of a star, a mass function $m(r)$ can be
defined (see, for example, Misner \& Sharp
\cite{art:MisnerSharp1964}). 
Buchdahl has derived inequalities limiting
the behavior of the mass  for solutions with a regular center
\begin{eqnarray}
  \label{eq:buch}
   m &\geq& \tfrac{4}{3}\pi r^3\rho\ ,\mathletter{a}\label{eq:buch1}\\
   m &\leq& \tfrac{4}{3}\pi r^3\rho_c\ ,\mathletter{b}\label{eq:buch2}\\
   0 &\geq& 9\left( \frac{m}{r} \right)^2 + 
            \frac{4m}{r}(6{\pi}r^2p - 1) + 4{\pi}r^2p(4{\pi}r^2p - 2)\ ,
            \mathletter{c}\label{eq:buch3} 
  \end{eqnarray}
where $\rho_c$ is the energy density at the center of the star (see,
for example, Buchdahl \cite{art:Buchdahl1959} and Hartle 
\cite{art:Hartle1978}). Alternatively, (\ref{eq:buch3}) can be written as
\begin{equation}
  \label{eq:buch3b}
   \frac{m}{r} \leq \tfrac{2}{9}
   \left( 1 - 6{\pi}r^2p +  \sqrt{1 + 6{\pi}r^2p} \right)\ .
\end{equation}
These inequalities are satisfied by all
regular solutions, but are equalities {\em only} for the incompressible
fluid. For such models, (\ref{eq:buch1}) and (\ref{eq:buch2})
characterize 
the regular solutions, while (\ref{eq:buch3}) corresponds to a certain
non-regular solution with a positive mass singularity.

A static spherically symmetric perfect 
fluid model has to be isolated, {\em i.e.}, it has to have a boundary at a 
finite radius, if it is to be considered as a star model.
At the radius where the pressure of the fluid vanishes, the 
interior solution can be matched with the static Schwarzschild solution
\begin{equation}\label{eq:schwa}
  ds^2 = -\left(1-\tfrac{2M}{r}\right)dt^2 +
  \frac{dr^2}{\left(1-\tfrac{2M}{r}\right)} + r^2d\Omega^2\ ,
\end{equation}
(see Schwarzschild \cite{art:Schwarzschild1916}).

If (\ref{eq:buch3b}) is evaluated at the surface of the star,
defined by $p=0$, we obtain
\begin{equation}
\label{eq:buchsurf}
 \tfrac{2M}{R} \leq \tfrac{8}{9}\ ,
\end{equation}
where $M$ is the mass and $R$ the radius of the star. 

The outline of the paper is as follows: 
Section \ref{sec:polytrope} constitutes the main part of the paper.
In this Section, the field equations are recast into two complementary
regular dynamical systems on compact state spaces.
Both systems are studied numerically and qualitatively. The results are 
compared and various physical implications are discussed.
For  example, it is shown that general relativistic 
models have finite radii when the polytropic index $n$ satisfies
$0\leq n \lesssim 3.339$ and infinite radii when $n\geq 5$. When
$3.339\lesssim n<5$, there exists
a 1-parameter set of models with finite radii and a finite number,
depending on $n$, with infinite radii.
The situation thus differs from the Newtonian case where
regular models have finite radii when
the polytropic index satisfies $0 \leq n < 5$, otherwise not.
In Section \ref{sec:general} we discuss how one can adapt the different 
approaches of this paper to facilitate studies of more general barotropic
equations of state. We conclude in Section \ref{sec:conclude} with some
remarks. In Appendix A, we indicate how the various variables were found.
In Appendix B we show the relationship between our variables and the 
so-called Lane-Emden variables.  

\vspace{0.7cm}

Throughout the paper, geometric units with $c=G=1$ are used, where $c$
is the speed of light and $G$ the gravitational constant. Roman
indices, $a,b,...=0,1,2,3$ denote spacetime indices.


\section{Dynamical systems and dynamical systems analysis}
\label{sec:polytrope}

We now present two complementary formulations adapted to
the polytropic equation of state (\ref{eq:polyeqstate}). 
In the first formulation, we introduce
dimensionless variables  
\begin{equation}
\label{eq:SyK}
\left\{\Sigma,K,y\right\}\ ,
\end{equation}
according to
\begin{equation}
  \label{eq:polyvarI}
  \tilde{N}^2 = y^2K\ , \quad {\phi}' = {\Sigma}y\ ,\quad
  r^2 = \frac{k^n P}{8\pi
  K}\left(\frac{1-y}{y}\right)^{1+n} \ , \quad y = \ydef\ ,
\end{equation}
where the prime denotes differentiation with respect to the
independent dimensionless spatial variable defined by the
above choice of $\tilde{N}$. The quantity $P$ is defined by
\begin{equation}\label{eq:PKS}
 P = 1-\Sigma^2-K\ ,
\end{equation}
and can be expressed as
\begin{equation}
\label{eq:Pdef}
 P = 8{\pi}Kr^2p\ .
\end{equation}
Thus, the assumption of a non-negative pressure implies $P\geq0$. The gauge, 
however, breaks down when $P=0$. From
(\ref{eq:polyvarI}), it follows that $K$ is
positive. We assume that the energy density $\rho$ and the pressure
$p$ are both non-negative. This, together with the defining
equation for $y$ in (\ref{eq:polyvarI}), implies that the variable $y$
satisfies $0 \leq y \leq 1$. 

For a non-negative polytropic index,
$n\geq0$, and for a positive constant $k$ in $p = k\rho^{1 + 1/n}$, 
the subset $y=0$ corresponds to the limit 
$\rho,p \rightarrow 0$, while the subset $y=1$ corresponds to the limit 
$\rho,p \rightarrow \infty$. However, one can also vary $k$.
Setting $k=0 \Rightarrow p=0$, while $k^{-1}=0 \Rightarrow \rho=0$.
These limits corresponds to $y=0$ and $y=1$, respectively,
as follows from the definition of $y$ in (\ref{eq:polyvarI}). 
The same ``equations of state''
($p=0$ or $\rho=0$) can be obtained from the linear equation of 
state, $p=(\gamma-1)\rho$; $p=0$ is obtained when $\gamma=1$ and $\rho=0$
is obtained when $\gamma \rightarrow \infty$.
Thus the subsets $y=0$ and $y=1$ can be viewed as state spaces
associated with the linear equation of state, $p=(\gamma-1)\rho$, in the 
limits $\gamma\rightarrow 1$ and $\gamma\rightarrow \infty$, respectively.

\begin{figure}[ht]
  \begin{center}
    \epsfig{figure=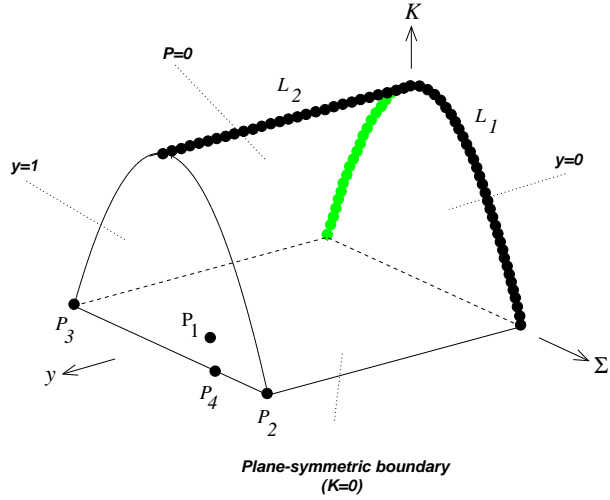, width=0.7\textwidth}
  \end{center}
  \caption{The state space for polytropic models, together with the 
different boundary subsets, in terms of the variables 
$\left\{\Sigma,y,K\right\}$.}
    \label{egghead}
 \end{figure}

Integrating (\ref{eq:dphidp}), leads to
\begin{equation}
 e^\phi = \alpha\left( 1 - y \right)^{1 + n}\ ,
\end{equation}
where $\alpha$ is a freely specifiable constant corresponding to the
freedom of scaling the time coordinate $t$ in the line element
(\ref{eq:ds2}). This, in turn, reflects the freedom in specifying the
value of the gravitational potential $\phi$ at some particular value
of $r$. Matching an interior solution with the exterior Schwarzschild
solution when $p=0$, however, fixes this constant
to $\alpha=\sqrt{1-\tfrac{2M}{r}}$.
It is worth noting the simple relationship between the variable $\Sigma$ and
the logarithmic derivative of the gravitational potential, 
\begin{equation}\label{eq:polyagrav}
 \dphi = \frac{\Sigma}{1-\Sigma}\ .
\end{equation}
In terms of the variables $\left\{\Sigma,K,u\right\}$, the gravitational 
field equations take the form
\begin{eqnarray}
\label{eq:SyKsyst}
  \Sigma' &=& -yK\Sigma + \tfrac12P\left(1+2y-4y\Sigma\right)\ ,
             \mathletter{a}\label{eq:polyeqaa}\\
  K' &=& 2y(\Sigma^2-P)K\ ,\mathletter{b}       \label{eq:polyeqab}\\
  y' &=& -(1 - \Gamma^{-1})y(1-y)\Sigma\ ,\mathletter{c}
             \label{eq:polyeqac} 
\end{eqnarray}
where $\Gamma=1+\tfrac{1}{n}$. It follows from
(\ref{eq:polyeqab}) and (\ref{eq:polyeqac}) that $K =0$ and
$y=0,1$ are invariant subsets. 
By differentiating (\ref{eq:Pdef}), we obtain
\begin{equation}
 P' = \left[ -\Sigma(1 + 2y + 2y\Sigma) + 2Ky\right]P\ ,
\end{equation}
which shows that $P=0$ is an invariant subset as well. All of these invariant
subsets constitute the boundary of the physical state space, and by
including them in the analysis we obtain a compact state space. 
For the $K=0$ subset, the 
remaining equations describe plane-symmetric polytropic models
(compare with the linear equation of state discussion 
in \cite{art:grstar_leq}),
and we therefore refer to 
this boundary subset as the plane-symmetric boundary. The state space
is shown in figure \ref{egghead}, along with the different
boundary subsets.

Differentiating $r$ in (\ref{eq:polyvarI}), yields
\begin{equation}
  \label{eq:rdefI}
  r' = (1 - \Sigma)yr\ ,
\end{equation}
which shows that $r$ is a monotone function for $y\neq0$, $r\neq 0$ and
$\Sigma\neq 1$. Together with monotone functions on the boundary subsets, 
this implies that all attractors are equilibrium  
points on the boundary subsets $y=0,1$ and $P=0$. The equilibrium
points of (\ref{eq:polyeqaa})-(\ref{eq:polyeqac}) and their 
corresponding eigenvalues are given in Table \ref{tab:original}. 
See also figure \ref{egghead}.

\begin{figure}[ht]
  \begin{center}
    \epsfig{figure=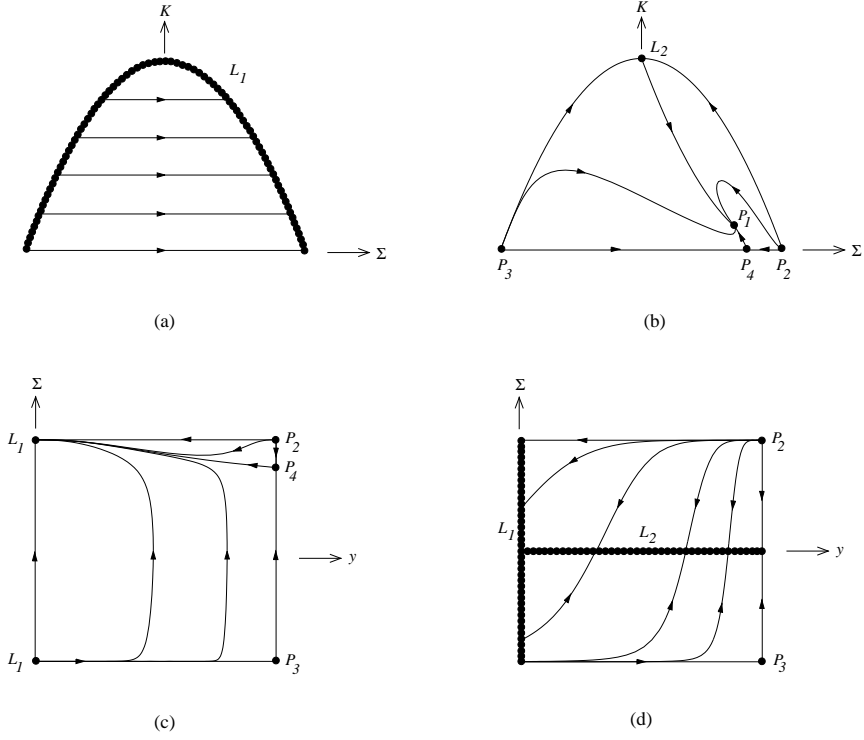, width=\textwidth}
    \caption{Orbits in the boundary subsets
      using the variables $\left\{\Sigma,K,y\right\}$ for (a) the
      boundary subset $y=0$, (b) the boundary subset $y=1$, (c) the
      plane-symmetric boundary $K=0$. (d) The $P=0$ subset
      projected onto the $K=0$ plane.} 
    \label{svampskalle}
    \end{center}
\end{figure}

\begin{table}[ht]
  \begin{center}
    \begin{tabular}{ccccc}
      \hline
      Eq point & $\Sigma$ & $K$ & $y$ & Eigenvalues \\ 
      \hline
      $L_1$ & $\Sigma_{\rm s}$ & $1-\Sigma_{\rm s}^2$ & 0 &
    $-\Sigma_{\rm s}\ , \ 0 \ , \  
      -\tfrac{\Sigma_{\rm s}}{1+n}$ \\
      $L_2$ & 0 & 1 & $y_{\rm c}$ & $0\ , \ -y_{\rm c}\ , \ 2y_{\rm
    c}$ \\ 
      $P_1$ & $\tfrac{2}{3}$ & $\tfrac{1}{9}$ & 1 & $-\tfrac{1}{3} \ ,
      \ -\tfrac{2}{3}\ , \ \tfrac{2}{3(1+n)}$ \\
      $P_2$ & 1 & 0 & 1 & $1 \ , \ 2\ , \ \tfrac{1}{1+n}$ \\
      $P_3$ & -1 & 0 & 1 & $7\ , \ 2 \ , \ -\tfrac{1}{1+n}$\\
      $P_4$ & $\tfrac{3}{4}$ & 0 & 1 & $-\tfrac{7}{8} \ , \
      \tfrac{1}{4} \ , \ \tfrac{3}{4(1+n)}$ \\
      \hline
    \end{tabular}
    \caption{The equilibrium points and their stability using the 
             variable set $\left\{\Sigma,K,y\right\}$.} 
    \label{tab:original}
  \end{center}
\end{table}
The solution structure of the boundary subsets is shown in figure
\ref{svampskalle}.

The equilibrium point $P_1$ is associated with the self-similar Tolman
solution, discussed in \cite{art:grstar_leq}, in the limit
$\gamma\rightarrow\infty$ when $p=(\gamma-1)\rho$. We therefore refer to 
this point as the Tolman point. 
A single orbit enters the interior state space from $P_1$ (it is associated 
with the single positive eigenvalue of $P_1$).  We refer to this orbit
as the Tolman orbit, although the corresponding exact solution is not known
(except in the incompressible fluid case). An approximation of this
solution near the center has been given by deFelice {\em et al} 
\cite{art:deFeliceetal1995}.
There originates a 2-parameter set of orbits into the interior state space
from the hyperbolic source $P_2$. These orbits correspond to solutions with a 
negative mass singularity. There are no interior orbits associated with the
hyperbolic saddle $P_3$. A 1-parameter set of orbits enters the interior 
state space from the hyperbolic saddle $P_4$. This point is associated with 
a self-similar plane symmetric solution discussed in \cite{art:grstar_leq} 
in the limit $\gamma\rightarrow\infty$ when $p=(\gamma-1)\rho$. The orbits
coming from this point are associated with a negative mass singularity.

The $\Sigma_{\rm s}<0$ part of the line $L_1$ constitutes a transversally
hyperbolic  
source from which a 2-parameter set of orbits originate. Solutions associated
with these orbits start out with a negative mass. Eventually,
for a sufficiently large radius $r$, all solutions starting out with negative 
mass acquire a positive mass, since the pressure is increasing
with $r$ when the mass $m(r)$ is negative and since the pressure and
energy density are assumed to be positive. The
$\Sigma_{\rm s} \geq 0$ part of the line $L_1$ constitutes the global sink. 
We note that the point $\Sigma_{\rm s}=0$ 
on $L_1$ is non-hyperbolic, with 
all three eigenvalues equal to zero. This makes it hard to analytically, as 
well as numerically, to resolve the dynamics in the vicinity of this point. 
One possibility to gain further information is to consider 
center-manifold theory, see Carr \cite{book:Carr1981} for an introduction 
to the subject. 
We will not pursue this possibility here. 
Instead we will rely on an indirect discussion concerning
the models radii and masses and on a second formulation, 
introduced below, which to a large extent resolves this problem. 

Each point in the equilibrium
set $L_2$ corresponds to the 
flat Minkowski geometry written on spherically symmetric form (compare
with the treatment in \cite{art:grstar_leq} where this solution was 
represented by a single equilibrium point). 
A single orbit, associated with the eigenvalue $2y_{\rm c}$, 
enters the interior state space from each point of $L_2$ when $y_c>0$. The 
1-parameter set of all such orbits form a separatrix surface
describing the regular subset of solutions, since the corresponding solutions 
all have regular centers.
The individual solutions are parametrized by $y_{\rm c}$,
which in turn is determined by the value of the central 
energy density, $\rho_{\rm c}$, 
and central pressure, $p_{\rm c}$, according to
$y_{\rm c}=\tfrac{p_{\rm c}}{p_{\rm c} + \rho_{\rm c}} =
\tfrac{k\rho_{\rm c}^{1/n}}{k\rho_{\rm c}^{1/n} + 1}$.
Demanding a causal star model, {\em i.e.}, a model where the 
velocity of sound of the fluid is less than or 
equal to the speed of light, yields the following inequality for the initial 
values $y_{\rm c}$ at the center of the star:
\begin{equation}
  \label{eq:causal}
  y_{\rm c} \leq \frac{n}{1+2n}\ .
\end{equation}

Where an orbit ends on $L_1$ is intimately connected with the solutions total 
mass and radius. The ratio between the mass function $m(r)$ and the radius 
$r$, is given by (as follows from the definition of $m$ given by
Misner \& Sharp \cite{art:MisnerSharp1964}) 
\begin{equation}\label{eq:mrKS}
 \frac{m}{r} = \frac{K-(1-\Sigma)^2}{2K}\ .
\end{equation}
It follows from the above equation that the static condition 
$\tfrac{m}{r} < \tfrac12$ is automatically satisfied in the interior 
state space. 
Note also that it follows that $\Sigma>0$ for solutions with positive mass.
Evaluating (\ref{eq:mrKS}) on $L_1$, 
where $K=K_{\rm s} = 1 - \Sigma_{\rm s}^2$, yields
\begin{equation}
 \label{eq:mrpolya}
 \frac{M}{R} = \frac{\Sigma_{\rm s}}{1 + \Sigma_{\rm s}}\ .
\end{equation}

A linear analysis shows that orbits ending at $\Sigma_{\rm s} > 0$ on $L_1$
correspond to solutions with finite radii and masses. 
Orbits ending at the non-hyperbolic point $\Sigma_{\rm s}=0$ on $L_1$
describe solutions with infinite radii and finite masses, or
masses that approach infinity slower than $r$. 

The Buchdahl inequalities limit the possible behavior of the regular subset.
Expressed in the present variables the first inequality, (\ref{eq:buch1}),
takes the form
\begin{equation}
 2y(3\Sigma(1-\Sigma) - P) \geq P\ ,
\end{equation}
where $P$ is given in (\ref{eq:PKS}).
For regular orbits, characterized by $y_c$, (\ref{eq:buch2}) yields
\begin{equation}
 K - (1-\Sigma)^2 \geq K P 
 \left( \frac{1-y}{y} \right)^{1+n} \left( \frac{y_c}{1-y_c} \right)^{1+n}\ .
\end{equation}
The third Buchdahl inequality, (\ref{eq:buch3}), leads to
\begin{equation}
 K \geq (1 - 2\Sigma)^2\ .
\end{equation}
On the surface this inequality results in $\Sigma_{\rm s} \leq \tfrac{4}{5}$.

\begin{figure}[ht]
  \begin{center}
    \epsfig{figure=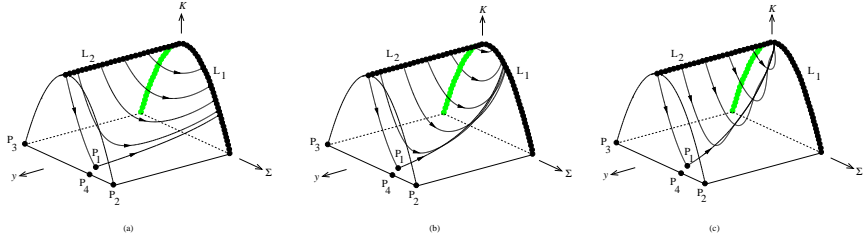, width=\textwidth}
    \caption{Orbits belonging to the regular subset
      using the variables $\left\{\Sigma,K,y\right\}$ for
      (a) $n=0$ (b) $0<n<5$ (c) $n\geq5$.} 
    \label{origreg}
  \end{center}
\end{figure}

In figure \ref{origreg}a,b,c some orbits in the regular
subset are shown, together with the important Tolman orbit.
As seen from the figures, 
the regular orbit on the $y=1$ boundary, connecting the $y=1$ end of
$L_2$ and the Tolman equilibrium point $P_1$, and the Tolman orbit
play important roles for understanding the regular
solutions, the latter acting not only as a boundary but also as a ``skeleton'' 
orbit for the regular set, when $n>0$. 
The situation strongly resembles that of 
the linear equation of state discussed in \cite{art:grstar_leq}.
This is not a coincidence. 
The regular orbit on the $y=1$ boundary and the Tolman orbit are completely 
analogous to the corresponding solutions in the linear equation of state 
case. As discussed 
earlier the $y=1$ set corresponds to the limit $\gamma \rightarrow \infty$
when $p=(\gamma-1)\rho$.
From the figures we also see that  as $n$ increases, the orbits are pushed
towards the point $\Sigma_{\rm s}=0$ on $L_1$
and that the maximum value of $\Sigma_{\rm s}$ for the regular solutions
is closely connected with the value of $\Sigma_{\rm s}$ for the Tolman orbit.

For the incompressible fluid case, $n=0$, the Tolman orbit
describes a ``simple'' lower boundary of the regular subset and gives an 
upper bound $\Sigma_{\rm s} =\tfrac{4}{5}$ 
for the regular solutions when they end 
at $L_1$, see figure \ref{origreg}a. This upper bound is just the
Buchdahl inequality (\ref{eq:buch3}), which holds for all equations of state. 
When $0<n<5$, the situation is more complicated. In this case some regular 
orbits spiral around the Tolman orbit and some regular orbits
end at larger $\Sigma_{\rm s}$-values on $L_1$ than the Tolman orbit. 
Nevertheless,
the largest $\Sigma_{\rm s}$-value is close to the one 
determined by the Tolman orbit
(for which $\Sigma_{\rm s} <\tfrac{4}{5}$ when $n\neq0$), 
see figure \ref{origreg}b. 
When $0\leq n\leq3$ all orbits end on $\Sigma_{\rm s} >0$; when
$3<n<5$ a 1-parameter set of orbits end on $\Sigma_{\rm s} >0$ 
and a finite number
at $\Sigma_{\rm s} =0$, as follows from the results obtained 
from the second formulation below.
When $n\geq5$, all regular orbits spiral around the Tolman orbit and they all 
end at $\Sigma_{\rm s} =0$, see figure \ref{origreg}c.

Combining the above discussion
with equation (\ref{eq:polyagrav}) tells us that
the behavior of the gravitational potential close to the surface 
of a regular solution (which might be located at infinity),
is to a considerable degree limited by the behavior of the gravitational
potential of the non-regular solution that corresponds to the Tolman orbit.
Moreover, this behavior is intimately connected
with the equation of state.

An advantage of the above formulation is the regularity of the equations which
makes the analysis of the center of the stars simple
(compare with Rendall \& Schmidt's analysis of singular differential equations
in \cite{art:RendallSchmidt1991}). 
In addition the above formulation works well when the energy density 
is large, even though one might be far from the center. 
The formulation also works well for any values of the energy density
when $n$ is small. Unfortunately, it is not well
suited to deal with the low energy density limit when $n$ is 
close to 3 or higher.

To facilitate a study of polytropic stars at low energy densities
for any non-negative value $n$, we now present another 
formulation of the polytropic equations. We again choose
\begin{equation}
 y=\ydef
\end{equation}
as a variable, but make a non-linear transformation of
$K$ and $\Sigma$ to a new set of dimensionless variables 
$U$ and $V$ (see Appendix A).
We first introduce the Newtonian homology invariants
$u,v$ (see, for example, Kimura \cite{art:Kimura1981})
\begin{equation}
 u = \frac{4{\pi}r^3\rho}{m}\ ,\quad v = \frac{m\rho}{rp}\ ,
\end{equation}
and then define the variables $U$ and $V$ according to
\begin{equation}\label{eq:UVdef}
  U = \frac{u}{1 + u} = \frac{4{\pi}r^2\rho}{4{\pi}r^2\rho + m/r}\ ,\quad
  V = \frac{v}{1 + v} = \frac{m/r}{m/r + p/\rho}\ .
\end{equation}
The assumptions of positive energy density, pressure, and mass lead
to the inequalities $0 < U < 1$ and $0 < V < 1$.
It follows that
  \begin{eqnarray}
    \label{eq:rphiUV}
  r^2 &=& \frac{k^nUV}{4\pi(1-U)(1-V)}\left(\frac{1-y}{y}\right)^{n-1}\
  , \mathletter{a} \label{eq:r2}\\ 
  e^\phi &=& \alpha\left( 1 - y \right)^{1 + n}\ . \mathletter{b}
  \end{eqnarray}

We now choose a new independent variable such that the function
$\tilde{N}$ takes the form
\begin{equation}
  \label{eq:lapseUV}
 \tilde{N}^2 = (1-y)^3(1-V)(1-U)^2Fr^{-2}\ ,\quad F =(1-V)(1-y) - 2yV\ ,
\end{equation}
where $r^2$ has been given in equation (\ref{eq:r2}).
We thereby incorporate the assumptions of positive energy density,
pressure, and mass in the gauge, which becomes ill-defined when these
requirements are not satisfied.

In terms of the variables $\left\{U,V,y\right\}$, the gravitational field 
equations, which are again completely regular, takes the form
  \begin{eqnarray}
  \label{eq:polyeqb}
    U' &=& U(1-U)\left[(1-y)(3-4U)F - \Gamma^{-1}G\right] 
           \label{eq:UVUeq} \ , \mathletter{a}\\ 
    V' &=& V(1-V)\left[(1-y)(2U-1)F + (1-\Gamma^{-1})G\right] 
           \label{eq:UVVeq}\ , \mathletter{b}\\ 
    y' &=& -(1-\Gamma^{-1})y(1-y)G \label{eq:UVyeq}\ , \mathletter{c}
  \end{eqnarray}
where
\begin{equation}
  G = V\left[ (1-U)(1-y) + yU \right]\ ,
\end{equation}
and where $\Gamma = 1 + 1/n$.
The prime denotes differentiation with respect to the independent
variable, defined by (\ref{eq:lapseUV}). By including the invariant
subsets $U=0,1,V=0,1,y=0,1$ of (\ref{eq:UVUeq})-(\ref{eq:UVyeq})
we obtain a compact
state space. These subsets can be interpreted in terms of
limits of $\tfrac{m}{r}$, $\tfrac{p}{\rho}=k\rho^{1/n}$
and $\rho r^2$.
The incorporation of the positive pressure, energy density and mass conditions
in the gauge has led to that we now have
invariant subsets corresponding to setting these quantities to zero.
From (\ref{eq:UVyeq}) it follows that $y$ is a monotonically
decreasing function, and all orbits in the interior of the state space  
approach the $y=0$ subset. The equilibrium points of
(\ref{eq:UVUeq})-(\ref{eq:UVyeq}) and their eigenvalues are given in
Table \ref{tab:UVcube}.   

Before continuing the discussion of the system
(\ref{eq:UVUeq})-(\ref{eq:UVyeq}), it is of interest to 
consider the Newtonian equations in terms of the variables
$\left\{U,V,y\right\}$. These are given by:
  \begin{eqnarray}
  \label{eq:polyeqbn}
  U' &=& U(1-U)\left[(3-4U)(1-V) - \Gamma^{-1}V(1-U)\right]\ , 
         \label{eq:Npolyeqa}\mathletter{a}\\
  V' &=& V(1-V)\left[(2U-1)(1-V) + (1 - \Gamma^{-1})V(1-U)\right]\ , 
         \label{eq:Npolyeqb} \mathletter{b}\\
  y' &=& -(1 - \Gamma^{-1}) y(1-y)V(1-U)\
  . \mathletter{c}\label{eq:Npolyeqc} 
  \end{eqnarray}
Note that the non-homologous variable $y$ is decoupled in the above
system, leaving a coupled 2-dimensional reduced homology invariant set of
equations. The reduced homology-invariant set of equations is identical
to the $y=0$ subset of the relativistic equations (\ref{eq:UVUeq}) 
and (\ref{eq:UVVeq}). We therefore refer to the $y=0$ subset
as the Newtonian subset. 
Taking the Newtonian limit of all the relativistic equations
(\ref{eq:UVUeq})-(\ref{eq:UVyeq}) thus corresponds to letting $y$ go to
zero (except in the equation for $y$ where one must be more careful). Hence
we obtain the Newtonian homology-invariant equations as the
``low pressure'' boundary of the relativistic state space.
It is worth noting that the projection of the Newtonian orbits onto
the subset $y=0$, 
coincides with the orbits on the same subset, because of the decoupling of 
the $y$-equation.

\begin{figure}[ht]
  \begin{center}
    \epsfig{figure=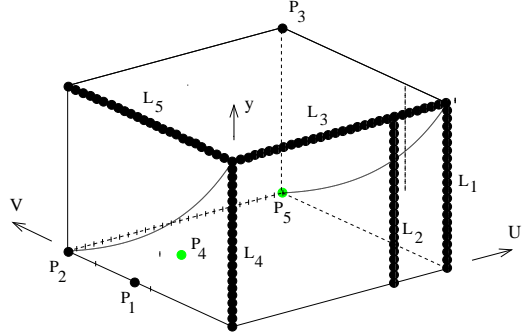, width=0.6\textwidth}
    \caption{The structure of the state space using the
      variables $\left\{U,V,y\right\}$. The curves indicated on the
      $U=0$ and 
      $U=1$ subsets are the intersections between these subsets and the
      ``static surface'' $(1+y)V-(1-y)=0$.}
    \label{pappskalle}
  \end{center}
\end{figure}

We now return to the relativistic case.
We are considering static models and thus the inequality
$\tfrac{2m}{r} < 1$ must be satisfied. Solving for $\tfrac{m}{r}$ in equation
(\ref{eq:UVdef}) leads to the expression
\begin{equation}\label{eq:mrUV}
  \frac{m}{r} = \left(\frac{V}{1-V}\right)\left(\frac{y}{1-y}\right)\ .
\end{equation}
Inserting this into $\tfrac{2m}{r} < 1$ yields the inequality
\begin{equation}
 \frac{1-y}{1+y} > V\ .
\end{equation}
The state space, together with the intersection of the
``static surface'' $(1+y)V - (1 - y) = 0$ with the $U=0$ and $U=1$ boundary 
subsets,
is depicted in figure \ref{pappskalle}. Note that the static surface
``forces'' the static solutions away from the line $L_5$ and the point
$P_3$, which thus are of little interest for the static solutions.

Expressed in the present variables, the first Buchdahl inequality 
(\ref{eq:buch1}) leads to
\begin{equation}\label{eq:buch1UV}
 U \leq \frac{3}{4}\ .
\end{equation}
This inequality ``forces'' the regular
subset away from $P_5$. 
Inserting equations
(\ref{eq:mrUV}) and 
$6\pi r^2 p = \tfrac{3}{2} UVy^2(1-U)^{-1}(1-V)^{-1}(1-y)^{-2}$
into (\ref{eq:buch2}) leads to a complicated expression, 
which we will not give.
For regular orbits characterized by $y_c$, equation (\ref{eq:buch2}) yields
\begin{equation}
 \left( \frac{y}{y_c} \right)  \left( \frac{1-y_c}{1-y} \right) \leq 
 \frac{1}{3} \left( \frac{U}{1-U} \right)\ .
\end{equation}

\begin{table}[ht]
  \begin{center}
    \begin{tabular}{cccccc}
      \hline
      Eq. point & $U$ & $V$ & $y$ & Eigenvalues & Restrictions \\
      \hline
      $L_1$ & 1 & 0 & $y_0$ & $(1-y_0)^2\ , \ (1-y_0)^2\ , \ 0$ & \\
      $L_2$ & $\tfrac{3}{4}$ & 0 & $y_c$ & $-\tfrac{3}{4}(1-y_c)^2\ , \
      \tfrac{1}{2}(1-y_c)^2\ , \ 0$ & \\
      $L_3$ & $U_0$ & 0 & 1 & $0 \ , \ 0 \ , \ 0$ & \\
      $L_4$ & 0 & 0 & $y_0$ & $3(1-y_0)^2\ , \ -(1-y_0)^2\ , \ 0$ & -- \\
      $L_5$ & 0 & $V_0$ & 1 & $0 \ , \ 0 \ , \ 0$ & -- \\
      $L_6$ & $U_0$ & 1 & 0 & $0\ , \ -(1-U_0)\ , \ -(1-U_0)$ & $n=0$ \\
      $P_1$ & 0 & $\tfrac{1+n}{2+n}$ & 0 & $-\tfrac{n-3}{2+n}\ , \
      \tfrac{1}{2+n} \ , \ -\tfrac{1}{2+n}$ & -- \\
      $P_2$ & 0 & 1 & 0 & $-\tfrac{n}{1+n} \ , \ -\tfrac{1}{1+n}\ , \
      -\tfrac{1}{1+n}$ & -- \\
      $P_3$ & 1 & 1 & 1 & $\tfrac{n}{1+n}\ , \ -\tfrac{1}{1+n}\ , \
      \tfrac{1}{1+n}$ & -- \\
      $P_4$ & $\tfrac{n-3}{2(n-2)}$ & $\tfrac{2(1+n)}{1+3n}$ & 0 &
      $\tfrac{-\lambda_3}{4}\left(5-n \pm
        \sqrt{a}\right)\ , \ \lambda_3$
      & $n>3$ \\
      $P_5$ & 1 & 1 & 0 & $0\ , \ 0 \ , \ 0$ & -- \\
    \hline
    \end{tabular}
  \end{center}
    \caption{The equilibrium points and their stability using the variables
$\left\{U,V,y\right\}$.
             Here $\lambda_3=-\tfrac{n-1}{(n-2)(1+3n)}$ and
             $a={1+22n-7n^2}$   
}
    \label{tab:UVcube}
\end{table}

A disadvantage with the present formulation is that the entire $y=1$
subset, in the $\left\{\Sigma,K,y\right\}$-formulation 
(see  figure \ref{svampskalle}(b)), 
has been ``crushed''
into the line $L_3$. We also note from Table \ref{tab:UVcube} that
this line is non-hyperbolic, with three zero eigenvalues. This means that
studying the dynamics in a neighborhood of this line can be quite
complicated. By relying on results from the 
$\left\{\Sigma, K, y\right\}$-formulation,  
we can circumvent most of these
problems. Although the previous formulation gives a better 
picture of what happens at large energy densities, the present
formulation gives a better picture of the low energy density
regime. The two formulations are thus complementary.

When discussing the solution structure using 
the $\left\{\Sigma,K,y\right\}$-formulation, we emphasized the
importance of certain special orbits, for example, the
Tolman orbit.
Thus it is important to identify this orbit in the present formulation.
The first step in order to do this is to identify where the Tolman point from 
which it originates is located. This point corresponds to
a point characterized by $U=(3+n)/2(2+n)$ and $V=0$ in the
equilibrium set $L_3$, as can be seen using the variable transformations
given in Appendix A. From these transformations one also sees that the
``plane-symmetric'' point $P_4$, 
in the $\left\{\Sigma, K, y\right\}$-formulation,
corresponds to the equilibrium point $U=0,V=0,y=1$, {\em i.e.}, the 
intersection
of the three lines $L_3,L_4$ and $L_5$. 
We know from the previous analysis that this latter point is associated with a
1-parameter set of 
solutions originating from it and that these solutions start out with negative
mass. Transforming from the
$\left\{\Sigma,K\right\}$-picture to the 
$\left\{U,V\right\}$-picture shows that the 1-parameter set 
of orbits starts from $U=0,V=0,y=1$ entering the 
``negative mass part'' of the state space outside the cube.
They eventually end at $L_1$ when the total mass becomes zero. This
line is just an artifact of the gauge choice. Using the previous formulation
one can continue the solutions through
$L_1$ into the the interior ``positive mass part'' of the cube.
The situation is analogous to that encountered in the diagonal homothetic
approach for self-similar spherically symmetric models when one matches 
the spatially self-similar part with the timelike self-similar part,
as described in Goliath {\em et al} 
\cite{art:Goliathetal1998a}, \cite{art:Goliathetal1998b}.
The 2-parameter set of orbits associated with $P_2$
in the $\left\{\Sigma, K, y\right\}$-formulation starts from $L_3$ and enters
into the negative mass regime of the state space outside of the cube, 
but eventually pass through $L_1$ into the interior state space, 
in the $\{U,V,y\}$-formulation.
The 2-parameter set of orbits associated with
the $\Sigma_{\rm s} <0$ part of $L_1$ 
(in the $\left\{\Sigma, K, y\right\}$-formulation) start from $L_1$
(in the $\{U,V,y\}$-formulation) entering the exterior state space 
and eventually pass through $L_1$ into the interior state space.

In the $\{U,V,y\}$-formulation,
all regular solutions start from the line $L_2$, and are again parametrized by
$y_c$. For $n=3$, the equilibrium point $P_4$ enters the
state space through the point $P_1$, changing the stability of $P_1$
from a source to a saddle. As $n$ increases, 
$P_4$ first changes from a node to a focus and then, at $n=5$, to a center
in the Newtonian subset. When $n>5$ $P_4$ is a local sink of the full state 
space. The point $P_4$ corresponds to a special singular Newtonian solution.

To be able to determine if an orbit corresponds to a model with finite 
mass or radius, it is useful to consider equation (\ref{eq:r2}) and
the following equations
\begin{eqnarray}
  \label{eq:polyeqbn}
  r' &=& (1 - y)(1 - U)Fr\ ,
         \label{eq:rpUV}\mathletter{a}\\
  m' &=& \frac{UV^2y^2Fm}{(1-V)^2(1-y)}\ , 
         \label{eq:mpUV} \mathletter{b}\\
  m^2 &=&
  \frac{k^nUV^3}{4\pi(1-U)(1-V)^3} \left(\frac{y}{1-y}\right)^{3-n} \
  , \label{eq:m2UV} \mathletter{c}
\end{eqnarray}
where $F =(1-V)(1-y) - 2yV$.

The radius of a model is definitely infinite if $r'$
does not approach zero as $\lambda\rightarrow +\infty$.
The corresponding statement holds for the mass, $m$, as well. 
The variable $y$ is monotonically decreasing and all orbits
end at the Newtonian boundary subset.
This, in conjunction with the stability of the equilibrium
points given in Table\ \ref{tab:UVcube}, implies the following:
{\em Only}
orbits that end at the hyperbolic sink $P_2$ (or the equilibrium set
$L_6$ for $n=0$) correspond to star models with finite radii. These 
models also have finite mass. If an
orbit ends at one of the other possible equilibrium points, $P_1$ or $P_4$, 
a model with infinite radius is obtained. If it ends at $P_1$ it 
yields a model with finite mass and if it ends at $P_4$ the corresponding 
model has infinite mass.  
This follows from a linear analysis of equations
(\ref{eq:r2}) and (\ref{eq:rpUV})-(\ref{eq:m2UV}).
Note that the above statements also hold for the Newtonian models, described by
the $y=0$ subset and the decoupled $y$-equation (\ref{eq:Npolyeqc}). 

We now consider the Newtonian subset $y=0$, since this subset is of great 
importance also for relativistic stars.

\subsection{The Newtonian subset}

Newtonian polytropic stars have been   
studied extensively in the literature. Chandrasekhar
\cite{book:Chandra1939} primarily used the Lane-Emden approach, while the
homology-invariant variables $\left\{u,v\right\}$ were used by, for
example, Kimura \cite{art:Kimura1981} and Horedt
\cite{art:Horedt1987}. Both Kimura as well as Horedt used
dynamical systems techniques in their investigations, but since they
did not use bounded variables, their state spaces were not
compact. 

\begin{figure}[ht]
  \begin{center}
    \epsfig{figure=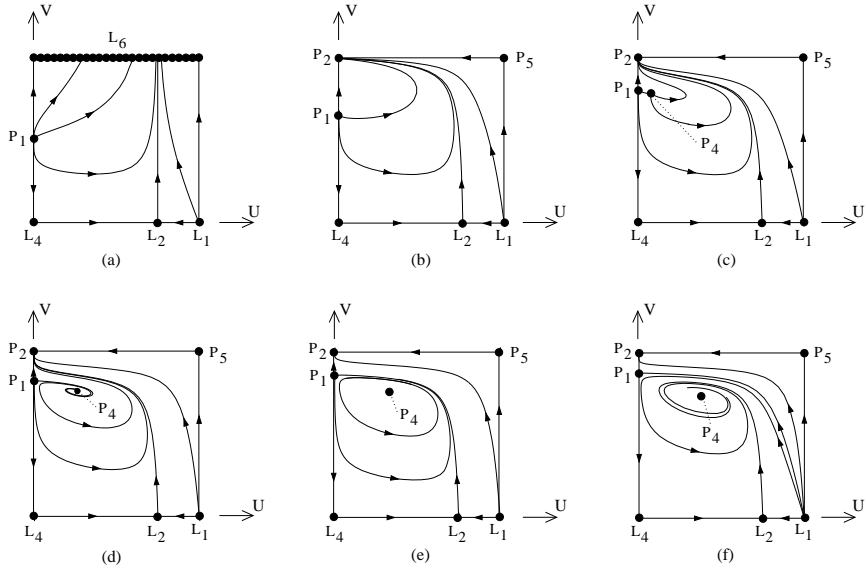, width=\textwidth}
    \caption{Orbits in the Newtonian subset $y=0$ in terms
      of the variables $\left\{U,V\right\}$ for (a) $n=0$, (b)
      $0<n\leq3$, (c) $3<n\leq(11+8\sqrt{2})/7$, (d)
      $(11+8\sqrt{2})/7<n<5$, (e) $n=5$, and (f) $n>5$.}
    \label{fig:newtonian}
  \end{center}
\end{figure}

From these previous analyses, however, it is known that regular
Newtonian solutions exist for all $n\geq0$, and that when $0\leq n<5$
they give rise to finite star models while
$n\geq5$ lead to infinite regular models.  
These results are easily obtained using the variables
$\left\{U,V\right\}$. The solution structure of the Newtonian
subset for different values of $n$ is shown in figure
\ref{fig:newtonian}. The orbit corresponding to the regular solution
(called the E-solution in the literature) starts at
$y=0$ on $L_2$, and we refer to it as the regular
Newtonian orbit. As indicated in the discussion above, the equilibrium
point where the orbit ends, depends on the value of $n$.
For the solvable incompressible fluid case $n=0$, the regular orbit is
characterized by $U=\tfrac{3}{4}$, and end on the corresponding point
on $L_6$. For $0<n<5$, it ends at
$P_2$. Thus, for these values of $n$, the radius and mass of the star model
are finite. We also note that for $0\leq n\leq3$ there exists a monotone 
function, 
\begin{equation}
Z = \frac{UV^3}{(1-U)(1-V)^3}\ , \label{eq:newtmono}
\end{equation}
which satisfies
\begin{equation}
\frac{Z'}{Z} = 2nU(1-V) + (3-n)V(1-U) .
\end{equation}

For $n=5$, however, the regular Newtonian orbit 
ends at $P_1$, and thus corresponds to a model with infinite
radius but with finite mass. Note that for $n=5$, the point $P_4$ is
a center. This is not just a feature of the linear analysis. The case $n=5$ 
is exactly solvable, see, for example, Chandrasekhar
\cite{book:Chandra1939}, so there exists an integral, which in our variables 
takes the form
\begin{equation}
   D(1-U)^{3/2}(1-V)^{5/2}= 6\sqrt{UV}H\ ,    
\end{equation}
with
\begin{equation}
   H = 2(1-U)(4V-3)^2 + 4UV(1-V) - 3(1-U)(1-V)^2\ ,
\end{equation}
where $D$ is a constant.
Setting $D=0$ yields the regular orbit. Negative values of $D$ parametrizes
the closed orbits around $P_4$. These orbits correspond to non-regular models 
with infinite radii. The orbits that end at $P_2$ are parametrized 
by the positive values of $D$. For $n>5$ all orbits end at $P_4$, giving
rise to models with infinite radii and masses.

\subsection{Regular relativistic stars}

The dynamics of the relativistic case is
much more complicated than in the Newtonian case.
In contrast to the Newtonian case, a projection of the orbits onto the 
plane $y=0$ does not coincide with the orbits in the $y=0$ subset,
since the equation for $y$ is not decoupled. 
The regular subset is a 1-parameter set of orbits
starting from $L_2$, defining a separatrix
surface in state space. In the interior of state space, 
the Tolman orbit constitutes the boundary of this separatrix surface.
Since $y$ is a monotone function, all orbits end at the Newtonian boundary.
We will now discuss the behavior of the orbits in the regular subset for 
different values of $n$.

\subsection{The incompressible fluid models}

The behavior of a relativistic incompressible fluid ($n=0$) is quite
reminiscent of the Newtonian case. All  orbits in the regular subset
lie in the plane $U=\tfrac{3}{4}$, and thus end at 
$U=\tfrac{3}{4}$ on $L_6$. The regular subset
projected on the Newtonian subset $y=0$ is shown in figure
\ref{fig:cubesubm}a.

We note that it is possible to solve the equations exactly for the
incompressible fluid, just as in the Newtonian case (see, for example,
Tooper \cite{art:Tooper1964}, who uses the relativistic Lane-Emden
equation).

\subsection{Models with $0< n \leq 3$}

For models with $0<n\leq3$, Table \ref{tab:UVcube} shows
that $P_2$ is a hyperbolic sink. The fact that $y$ is a monotone
function that pushes all orbits down to the Newtonian subset, in
conjunction with the existence of the monotone function
(\ref{eq:newtmono}) in the Newtonian subset for $0\leq n \leq 3$,
implies that $P_2$ is, in fact, the global sink and {\it all} orbits 
end at this point. Therefore the radii and masses of all regular 
stars are finite when $0<n\leq3$. 
This is consistent with a theorem given by Makino \cite{art:Makino1998}, 
which in the present context states that the radius of a regular model is 
finite when $1<n<3$ (see Theorem 1, p 60 in \cite{art:Makino1998}).

\begin{figure}[ht]
  \begin{center}
    \epsfig{figure=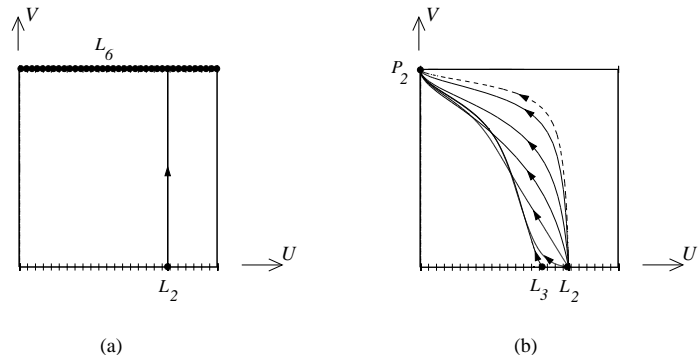, width=0.8 \textwidth}
    \caption{Orbits belonging to the regular subset  and the Tolman orbit
      using the variables $\left\{U,V,y\right\}$ for (a) the
      incompressible fluid, $n=0$, and (b) $0<n\leq3$. The Newtonian 
      orbit corresponds to the dashed line.}
    \label{fig:cubesubm}
  \end{center}
\end{figure}

The regular subset projected onto the plane
$y=0$ is shown in figure \ref{fig:cubesubm}b, where the regular
Newtonian orbit is shown as a dashed line. Also shown is the Tolman orbit, 
starting from $L_3$. Orbits characterized by
an initial value of $y_c$ close to unity, start out close to $L_3$, 
and then follow the Tolman orbit closely. Note that the separatrix surface 
formed by the regular subset of orbits folds over itself.   

\subsection{Models with $3< n < 5$}

For $3<n<5$, the point $P_2$ is still the only sink, 
and the majority of orbits will end at $P_2$. Hence most regular orbits 
are expected to end there too, but there are other
possibilities for where the orbits of the regular subset can end. There
exists a 1-parameter set of
orbits that end at $P_1$, and thus gives rise to a separatrix
surface in state space. 
The corresponding solutions have finite masses but infinite radii.
The boundary of the  
separatrix surface associated with $P_1$ is described by an orbit in the
$U=0$ submanifold, an orbit in the
Newtonian subset $y=0$ connecting the point $P_1$ with $P_4$, and
a relativistic orbit ending at $P_4$. The solution corresponding to this
latter orbit have both infinite mass and radius.
If the ``regular'' separatrix surface
intersects this separatrix surface or its boundary orbit 
in the interior state space,
there will be an orbit, ending at $P_1$ or at $P_4$, respectively, instead 
of $P_2$. The two surfaces could, of course, intersect each other several 
times, resulting in several infinite solutions. What actually happens 
depends on the value of $n$, as discussed below.

We have been forced to rely on numerical simulations in order to
decide where the orbits of the regular subset actually end.
To systematically explore if there are regular orbits that end at 
$P_1$ or $P_4$, we proceed as follows. 
We first note that when $3<n<5$, there always exist a constant $c_1>0$,
such that all orbits with $y_c<c_1$ end at $P_2$. This follows from our 
numerical investigation, but is also a consequence of Theorem 4 on p 994
in Rendall \& Schmidt \cite{art:RendallSchmidt1991}
(see also Theorem 2 p 64 in  Makino \cite{art:Makino1998}), which states
the following in the present context. Any relativistic polytropic model with
$1<n<5$ and central density $\rho_{\rm c}$ has finite radius if $k$ is 
sufficiently small. Since 
$y_{\rm c}=\tfrac{p_{\rm c}}{p_{\rm c} + \rho_{\rm c}} = 
\tfrac{k\rho_{\rm c}^{1/n}}{k\rho_{\rm c}^{1/n} + 1}$, this means that models 
with sufficiently small $y_{\rm c}$ lead to finite stars. 
Since $y$ is a monotone function, the intersections of the 
separatrix surfaces with a plane $y=c_2$ are curves. Hence, an intersection
of the separatrix surfaces corresponds to an intersection of the two 
corresponding curves in the plane $y=c_2>c_1$. To visualize the intersections
of the two separatrix surfaces for a given $n$, we thus 
first ensure, numerically,
that all orbits with $y_c\leq c_2$, where $c_2>0$ is chosen to be some 
convenient value, end at $P_2$. We then numerically calculate the separatrix 
surfaces and plot their intersection with the plane $y=c_2$. 

\begin{figure}[ht]
  \begin{center}
    \epsfig{figure=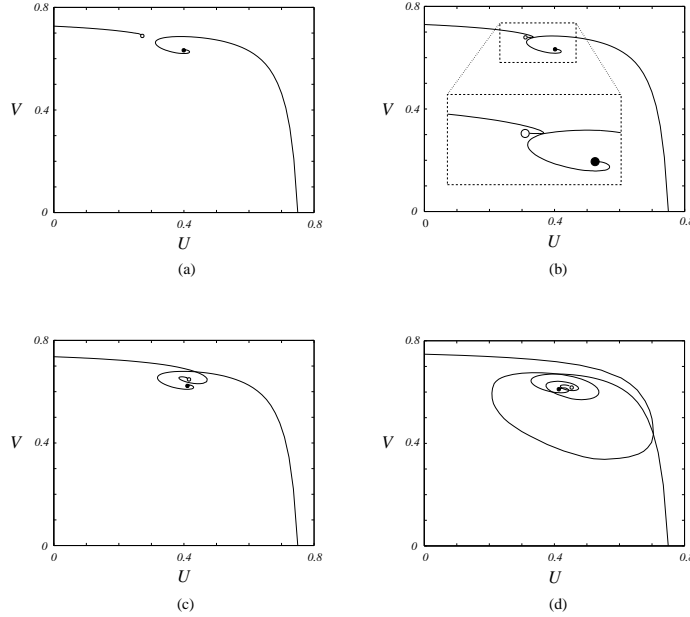, width=0.8 \textwidth}
    \caption{The intersection of regular subset and the separatrix surface 
ending at the equilibrium point $P_1$ in the plane $y=0.1$, for (a)
    $n=3.1$, (b) $n\approx 3.339$,   
(c) $n=3.6$, and (d) $n=4.1$.}
    \label{fig:smiley}
  \end{center}
\end{figure}

Numerical simulations indicate that all regular orbits
end at $P_2$ for $3 < n\lesssim 3.339$, and all
regular models are thus finite for this range of $n$, 
see figure \ref{fig:smiley}a.
The first time the two separatrix surfaces intersect each other is when
$n\approx 3.339$, see figure \ref{fig:smiley}b.
Thus there is a single regular orbit ending at $P_1$, 
corresponding to a solution with finite mass and infinite radius. 
All other regular 
orbits end at $P_2$, giving rise to models with finite radii and masses. 
When $n$ is larger than the above value there is always at least one 
separatrix intersection with a ``$P_1$-orbit'' and there is at least
one solution with finite mass and infinite radius. 

\begin{figure}[ht]
  \begin{center}
    \epsfig{figure=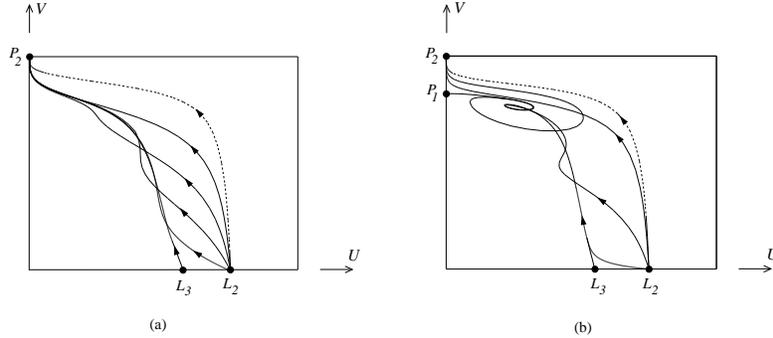, width=0.9 \textwidth}
    \caption{Orbits belonging to the regular subset and the Tolman
    orbit using the variables $\left\{U,V,y\right\}$, for (a) $3< n
    \lesssim 3.339$ and (b) $3.339\lesssim n <5$.}
    \label{fig:n3to5}
  \end{center}
\end{figure}

In figure \ref{fig:smiley}c the intersection between the
two separatrix surfaces for $n=3.6$ with the plane $y=c_2=0.1$ is shown. 
The single point of intersection of the two curves shows that there exists 
a single infinite solution. 
For $n=4.1$, the result is shown in figure \ref{fig:smiley}d. We note that
there are several infinite solutions, but they are all isolated, and this is 
typical. Intersections always seem to take place at isolated
points. The behavior of the regular subset projected onto the
Newtonian subset is shown in figure \ref{fig:n3to5}a for $3 < n
\lesssim 3.339$ and in figure \ref{fig:n3to5}b for $3.339 \lesssim n
<5$. One should note that as $n$ approaches 5, the behavior of the
regular subset becomes more and more complicated. 

The boundary orbit of the $P_1$-separatrix surface, corresponding to
the $P_4$ orbit, may also intersect the regular separatrix
surface. This happens when $n\approx3.357$ and $n\approx4.414$. In
these cases one obtains solutions with infinite masses and radii.  

The Tolman orbit, {\em i.e.}, the boundary of the regular subset, also
intersects the $P_1$-separatrix surface. This happens at an increasing
rate as $n$ approaches 5. The first five values of $n$ when this
happens are $3.673, 3.939, 4.105, 4.221$ and $4.309$. 

The lowest initial value of $y_c$ leading to an infinite model
decreases towards zero when $n$ increases towards 5.
For $n=5$ it coincides with $y_c=0$.
The relativistic 3-dimensional state space thus
sheds light on the appearance of an infinite Newtonian star for
$n=5$. 

It is worth noting that the appearance of
infinite solutions is a source of considerable problems in the
Lane-Emden approach. 

\subsection{Models with $n=5$}
\begin{figure}[ht]
  \begin{center}
    \epsfig{figure=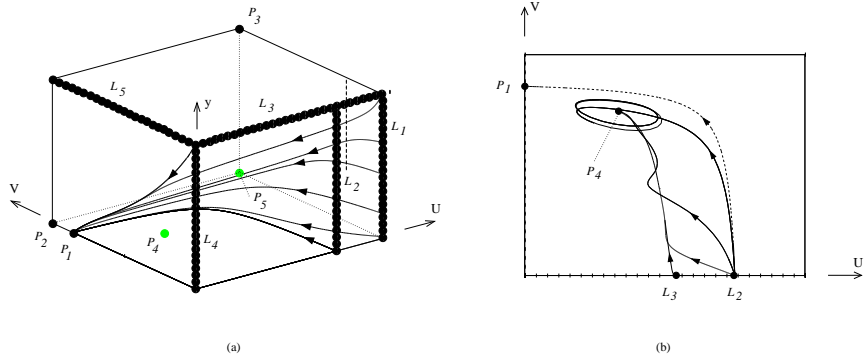, width=\textwidth}
    \caption{(a) The behavior of the separatrix surface
      ending at the equilibrium point $P_1$ that excludes any finite
      polytropic stars with $n=5$. (b) Orbits in the regular 
      subset and the Tolman orbit projected onto the Newtonian subset $y=0$.}
    \label{fig:n5sepsurf}
  \end{center}
\end{figure}

When $n=5$, the separatrix surface that ends at the equilibrium point $P_1$
completely encloses the regular subset of orbits. This makes it impossible
for any regular orbit to end at $P_2$. The Newtonian regular orbit
is the lower boundary of the separatrix surface and thus end at $P_1$,
while all relativistic regular orbits are forced towards $P_4$ and the 
closed orbits surrounding $P_4$, in the Newtonian subset $y=0$. 
All relativistic regular models therefore have infinite radii and masses.
The enclosing separatrix surface is shown in figure
\ref{fig:n5sepsurf}a, and the projection of the regular subset 
and the Tolman orbit onto the Newtonian subset $y=0$ is shown in 
figure \ref{fig:n5sepsurf}b.  

\subsection{Models with $n>5$}

\begin{figure}[ht]
  \begin{center}
    \epsfig{figure=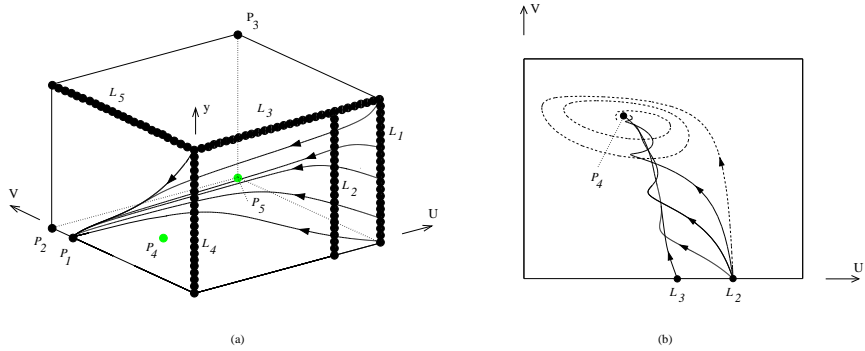, width=\textwidth}
  \end{center}
    \caption{(a) The behavior of the separatrix surface
      ending at the equilibrium point $P_1$ that excludes any finite
      polytropic stars with $n>5$. (b) The orbits of the regular subset 
      and the Tolman orbit projected onto the Newtonian subset $y=0$.}
    \label{fig:n5gsepsurf}
\end{figure}

When $n>5$, the situation is similar to the case $n=5$. The
separatrix surface associated with $P_1$
now encloses {\em all} regular orbits (including the
regular Newtonian orbit) and all regular
orbits end at the sink $P_4$. 
Hence all polytropic models with $n>5$, Newtonian and relativistic,
have infinite radii and masses.
The enclosing separatrix surface associated with $P_1$ is shown in figure
\ref{fig:n5gsepsurf}a, and the projection of the regular subset and the
Tolman orbit onto
$y=0$ is shown in figure \ref{fig:n5gsepsurf}b.

\section{Comments on the general barotropic
  equation of state} \label{sec:general}

In this section we address
the issue whether it is possible to use, or at least modify, the above
formulations, in order to understand models with other
barotropic equations of state. 

In general, a barotropic equation of state, $p=p(\rho)$, leads to that
\begin{equation}
 \Gamma = \frac{d\ln p}{d\ln \rho}\ 
\end{equation}
is a function of $y$. 
For a given $\Gamma(y)$, one can parametrize the equation of state
in terms of $y$ as follows:
\begin{eqnarray}
 \rho &=& \rho_0 \exp\left(\int \frac{dy}{({\Gamma}(y)
 -1)y(1-y)}\right)\ , \mathletter{a}\\ 
 p &=& p_0\exp 
 \left(\int \frac{{\Gamma}(y) dy}{({\Gamma}(y) -1)y(1-y)}\right)\
 . \mathletter{c} 
\end{eqnarray}
The parametrization, however, might
break down, but in such cases one can in principle
obtain the equation of state in
a piece-wise manner, 
except when $\Gamma =1$ everywhere. This latter situation
corresponds to a linear scale-invariant equation
of state, $p= (\gamma-1)\rho$, easily treated with other methods, see, for 
example \cite{art:grstar_leq}.

To obtain a formulation for non-polytropic equations of state,
one has to replace the constant $\Gamma$ in the equations in the previous 
section with the function $\Gamma (y)$. 
The functional form of the factor $1-\Gamma^{-1}$ for some common
equations of state is given in Table \ref{tab:eqofstate}.

\begin{table}
   \begin{tabular}{llc}
   {Type} & {Eq.\ of state} & $1-\Gamma^{-1}$ \\
  \hline 
  Incompressible & $\rho=\mathrm{const}$ & 1 \\
  Scale-invariant & $p \propto \rho$ & 0 \\
  Polytrope & $p \propto \rho^{1+1/n}$ & $(1+n)^{-1}$ \\
  Linear & $\rho = (\eta-1)p + \rho_0$ & $(1-\eta y)/(1-y)$ \\
  Relativistic polytrope & $p=k\rho_m^{\bar{\Gamma}}\ , \quad \rho=\rho_m + 
	p/(\bar{\Gamma}-1)$ & $(1-\bar{\eta} y)/[\bar{\eta}(1-y)]$
\end{tabular}
\caption{The functional form of $1-\Gamma^{-1}$ 
         for some common equations of state. Here $\rho_m$ is the rest mass
         density and ${\bar\eta}= {\bar\Gamma}/(1-{\bar\Gamma})$.}
\label{tab:eqofstate}
\end{table}

It should be pointed out that in general $y$ is not a monotone
single-valued function of $p$. 
Several values of $p$ and $\rho$ might correspond to
the same value of $y$. This can be visualized as follows: First note that 
there is a one-to-one correspondence between $y$ and
$\tfrac{\rho}{p}$ since $y=\tfrac{1}{1 + (\rho/p)}$.
Then plot the equation of state in a ${\rho}$-$p$ diagram. The slope
of a straight line from the origin determines
$\tfrac{\rho}{p}$ and therefore $y$. From this construction one can
deduce the functional properties of $y$ as a function of $p$ or
$\rho$. If a straight line intersects the equation of state curve
several times, then several values of $p$ and $\rho$ correspond to
the same value of $y$. 
This feature is of course encoded in the above equations.
If this situation occurs, then $y$ is not a monotone function and 
$1-\Gamma(y)^{-1}$ 
becomes zero for one or several values of $y$ in the interval
$0 < y < 1$. For these values of $y$ one obtains invariant subsets
described by the scale-invariant linear equation of state.
If there are only a few zeroes of $1-\Gamma^{-1}$, one may
match orbits and obtain the complete solution.

It is only possible to obtain $\Gamma$ as a function of $y$ explicitly 
for a limited set of equations
of state. Because of these difficulties, a better strategy for
most equations of state,
particularly if $1-\Gamma^{-1}$ has several zeroes, is to change from
$y$ to another variable $z=z(y)$, that is a monotone function of $p$, 
and treat $y$ as a function of $z$ in the equations and if
necessary choose a new independent variable. 
Simple examples of new variables are, for example, $p$,
some simple function of $e^\phi$ (like the redshift), or, if one
wants a bounded variable, for example, $z=\tfrac{p}{a^2 + p}$,
$z=\tfrac{e^\phi}{a^2 + e^\phi}$,
where $a^2$ is some
arbitrary constant with suitable dimension, but preferably
associated with the equation of state of interest.
It might happen that one obtains new boundaries replacing, for example, 
the $y=1$ boundary
for the physical interesting part. This is exemplified by the
linear equation of state, for which $1-\Gamma^{-1}=\tfrac{1 - \eta
y}{1-y}$ and the relativistic polytrope, for which
$1-\Gamma^{-1}=(1-\bar{\eta}y)/\bar{\eta}(1-y)$. 
In these cases one obtains a boundary at $y = \tfrac{1}{\eta}$ and $y
= \tfrac{1}{\bar{\eta}}$ respectively, instead of at $y=1$. Note that for
the system to be useful, one requires 
that the equations are differentiable of at least order
$C^1$ on the entire physical state space and its boundaries.

To exemplify that one can use the polytropic formulations for 
other equations of state, we will first
consider the linear equation of state
and make comparisons with the approach described in \cite{art:grstar_leq}.
In the $\{\Sigma,K,y\}$-formulation one only needs to replace $1-y$ with 
$1-\eta y$ in the equation for $y$. 
Hence $y = \eta^{-1} < 1$ replaces the boundary subset $y=1$ and
the physically interesting region is characterized by $y \leq \eta^{-1}\leq1$.
The resulting dynamical system is completely
regular on the physical state space and its boundaries. 

In \cite{art:grstar_leq} the state space
allowed orbits corresponding to solutions with negative pressure.
This is not the case now. We have incorporated the positive
pressure condition in the gauge and we therefore now have an
invariant subset corresponding to setting $p$ to zero.
Instead of ending at a non-invariant surface of zero pressure,
as in \cite{art:grstar_leq}, the regular orbits now end at 
$\Sigma_{\rm s}s>0$ on $L_1$, since all regular solutions have finite radii 
and
masses, see \cite{art:grstar_leq}. 
In addition, all regular solutions now start from a line of equilibrium 
points $L_2$, while in \cite{art:grstar_leq} they all started from a single
isolated equilibrium point.

For a linear equation of state the $\{U,V,y\}$-formulation yields the following
equations:
  \begin{eqnarray}
  \label{eq:genlinj}
  U' &=& U(1-U)\left[(1-y)(3-4U)F - \Gamma^{-1}G\right]\ ,
  \mathletter{a}\\ 
  V' &=& V(1-V)\left[(1-y)(2U-1)F +
  (1-\Gamma^{-1})G\right]\mathletter{b}\ , \\ 
  y' &=& -y(1-{\eta}y)G\ , \mathletter{c}
  \end{eqnarray}
where 
  \begin{eqnarray}
  \label{eq:FG}
  F &=& (1-V)(1-y) - 2yV\ , \mathletter{a}\\ 
  G &=& V[(1-U)(1-y) + yU]\ , \mathletter{b}
  \end{eqnarray}
and where
\begin{equation}
 1-\Gamma^{-1}=\frac{1 - \eta y}{1-y}\ .
\end{equation}
The expression $(1-y)^{-1}$ occurs as a factor in the above dynamical
system, except when $\eta=1$ and $\Gamma^{-1}=0$. This case corresponds
to an incompressible fluid and can thus
be obtained by setting $n=0$ in the polytropic equation of state.
For $\eta > 1$, $y = \eta^{-1} < 1$ replaces $y=1$ as the boundary
subset of the physically interesting region. Thus
$y \leq \eta^{-1}<1$, and this implies that $1-y>0$.
Hence the above system is completely
regular on the physical state space and its boundaries. 
If one is so inclined, one can change the independent variable
with a factor of $(1-y)$ so that one obtains polynomial equations.
Note that $y$ is a monotone function in the physically interesting region
for this case and thus there is no need to 
change to another variable $z=z(y)$.

Solutions that start with negative mass enter the interior state space
through a line of equilibrium points, which did not exist in the 
linear equation of state treatment in \cite{art:grstar_leq}. 
In addition, all regular solutions start from a line of equilibrium 
points, while they all started from a single isolated
equilibrium point in \cite{art:grstar_leq}. 
Moreover, instead of ending at a surface of vanishing 
pressure as in \cite{art:grstar_leq}, 
they all end at the equilibrium point $P_2$ on the $y=0$ subset. 

The pictures are thus quite different, but one can
extract all interesting physical information about, for example, the regular 
solutions from any of the above formulations. 
However, note that it is easier to extract information
about total masses and radii using the formulation given in
\cite{art:grstar_leq}.  

The relativistic polytropic equation of state
\begin{equation}
  p = k\rho_m^{\bar{\Gamma}}\ , \quad 
 \rho = \rho_m + \frac{p}{\bar{\Gamma}-1}\ , 
\end{equation}
where $\rho_m$ is the rest mass density and $\rho$ the energy 
density,  provides another example where the previous formulations 
works well. For this equation of state the polytropic
exponent ${\bar\Gamma}$ coincides with the adiabatic index $\Gamma_a$
\begin{equation}
  \Gamma_a = \left(\frac{\rho+p}{p}\right)\frac{dp}{d\rho}\ , 
\end{equation}
which is not the case for $\Gamma$ in $p=k\rho^\Gamma$.
This equation of state asymptotically approach the linear equation of
state $p = ({\bar\Gamma} - 1)\rho$ when $p,\rho\rightarrow \infty$ and the
polytropic equation of state $p=k\rho^{\bar{\Gamma}}$ when 
$p,\rho\rightarrow 0$. Hence, the $y=1$ boundary is 
replaced with a boundary $y=\tfrac{1}{\bar\eta}$, described by a
scale-invariant linear equation of state. The $y=0$ boundary is just
the Newtonian polytropic boundary discussed previously.


\section{Concluding remarks} \label{sec:conclude}

We have obtained two complementary formulations with
regularized equations on compact state spaces for
polytropic equations of state. This has allowed us to
obtain a global picture of the solution space of these models. 
The two approaches revealed that certain solutions play an important
role for the remaining solution structure. 

The first approach 
(the $\{\Sigma,K,y\}$-formulation) showed that the regular scale-invariant
solution, the self-similar Tolman solution, and a special non-regular solution
(corresponding to the Tolman orbit) to a large extent determine the
high energy density behavior of the regular solutions. Note that the
non-regular Tolman solution is scale-invariant towards the
origin. An approximate expressions for the metric coefficients for this 
solution has been given by deFelice {\em et al} \cite{art:deFeliceetal1995}. 

The second approach (the $\{U,V,y\}$-formulation)
revealed that when $n>3$ there exists another special solution 
(the one ending at $P_4$) of great importance for the remaining solution
structure. When $3<n<5$ other solutions spiral around this ``skeleton'' 
solution. When $n>5$, the regular solutions approach this solution
and end at its endpoint $P_4$. This point is associated with a 
non-regular Newtonian solution, which describes the behavior of
the regular solutions close to the surface when $n>5$. 

The dynamical behavior is particularly complicated 
when $3<n<5$. It is interesting to compare with
Theorem 4 p 994 in Rendall \& Schmidt 
\cite{art:RendallSchmidt1991}, which 
in the present context states that regular
relativistic polytropic models with $1<n<5$ have finite radii if 
$y_{\rm c}$ is sufficiently small, and Theorem 1 p 60 in Makino
\cite{art:Makino1998}, which in the present context states that 
when $1<n<3$, the radius of a regular model is finite. 
Makino also comments that
the restriction of Rendall \& Schmidt may be inevitable when $3\leq n<5$
(p 62 in Makino \cite{art:Makino1998}). For polytropes we have numerically
shown that {\it all} regular models are finite
when $n\lesssim 3.339$. Thus, in this polytropic index range the restriction
is {\it not} necessary. However, our investigation shows that for larger 
values of $n$ it is. Our work shows that what is actually happening 
when $3<n<5$ is determined in the relativistic regime and not the Newtonian
regime, although this latter regime determines a finite number of 
possibilities. Considering the complicated dynamical behavior, it is hard to 
see how one could obtain our results except by numerics. Perhaps one
could gain further insights by going beyond the static and spherical
assumptions. For example, the behavior might be partly related to  
the stability features of the models.

We have shown that the present methods can be generalized to other equations 
of state. The state spaces in such cases are also 3-dimensional and this will
make it possible to visualize the entire solution space for such equations of 
state as well. The first $\{\Sigma,K,y\}$-formulation should be useful
when an equation of state behaves linearly or polytropically when the
energy density goes to infinity. 
For such equations of state one could expect that
the regular scale-invariant solution, the self-similar Tolman solution, 
and a special non-regular solution
(the Tolman orbit), determine, to a large extent, the
high energy density behavior of the regular solutions.
The second $\{U,V,y\}$-formulation should be useful for equations of
state that behaves polytropically when the energy density goes to zero.
For such cases the low energy density regime should be described by the 
Newtonian boundary. An interesting question is what happens with the radius 
of a regular model when one has an equation of state that approaches a 
polytrope with index $3<n<5$ asymptotically when $p, \rho\rightarrow0$. It is
probably determined in the relativistic regime, which depends on the actual 
equation of state.  
The relativistic polytropic equation of state nicely illustrates
the above discussion. Since this equation of state is of considerable 
physical interest, we will treat this case in more detail in a future paper.

We think that the present type of approach nicely complements those used by 
Rendall \& Schmidt \cite{art:RendallSchmidt1991} and Makino 
\cite{art:Makino1998} and that they together should constitute useful 
tools for further explorations of relativistic star models.


\appendix{A}

In this Appendix we will show how the different sets of variables can
be derived. Starting with the line element in the form
\begin{equation}
 ds^2 = -{\rm e}^{2\phi}dt^2 +
 d\ell^2 + e^{2\psi - 2\phi} d\Omega^2\ ,
\end{equation}
and introducing variables according to 
\begin{equation}
  \theta = \dot{\psi}\ ,\quad \sigma = \dot{\theta}\ ,\quad
  B = e^{\phi -\psi}\ , 
\end{equation}
where a dot denotes differentiation with respect to $\ell$,
leads to the following expressions for the gravitational field
equations
  \begin{eqnarray}
  \label{eq:thetaa}
  \dot{\theta} &=& -2\theta^2 + \theta\sigma + B^2 + 16{\pi}p\ ,
  \mathletter{a}\\ 
  \dot{\sigma} &=& -2\theta\sigma + \sigma^2 + 4{\pi}(\rho + 3p)\
  , \mathletter{b}\\ 
  \dot{B} &=& (-\theta + \sigma)B\ , \mathletter{c}\\
  8{\pi}p &=& \theta^2 - \sigma^2 - B^2\ . \mathletter{d}
  \end{eqnarray}
This system looks very much like systems in spatially
homogeneous cosmology (see, for example, Wainwright \& Ellis
\cite{book:WainwrightEllis1997}), even 
though the physical interpretation is quite different.
Hence one can import ideas from treatments of spatially homogeneous
models to the present situation.

To obtain the $\left\{\Sigma,K,y\right\}$-formulation, 
we use similar variables to the so-called expansion-normalized variables
used frequently in spatially 
homogeneous cosmology (see Wainwright \& Ellis 
\cite{book:WainwrightEllis1997}), 
and a dimensionless matter variable $y$ according to
\begin{equation}
  \Sigma = \frac{\sigma}{\theta}\ , \quad
  K = \frac{B^2}{\theta^2}\ , \quad
  y = \frac{p}{p + \rho}\ ,\quad
  P = \frac{8{\pi}p}{\theta^2}\ .
\end{equation}
The new independent variable is determined by
\begin{equation}
 \frac{d\ell}{d\lambda} = \frac{y}{\theta}\ .
\end{equation}

To obtain the $\left\{U,V,y\right\}$-formulation \ref{sec:polytrope}, we 
start with the Tolman-Oppenheimer-Volkoff 
approach 
  \begin{eqnarray}
  \label{eq:TOV}
 \frac{dp}{dr} &=& -\frac{(p + \rho)(m + 4{\pi}r^3p)}{r(r - 2m)}\ ,
 \mathletter{a} \\
 \frac{dm}{dr} &=& 4{\pi}r^2\rho\ , \mathletter{b}
  \end{eqnarray}
(see, for example, \cite{book:gravitation}).
We then introduce the variables $\{U,V,y\}$ in terms of
$r,m,p$, and $\rho$, according to the definitions
(\ref{eq:UVdef}) and (\ref{eq:polyvarI}), and
a new independent variable defined by
\begin{equation}
 \frac{dr}{rd\lambda} = (1 - y)(1 - U)F\ ,\quad F =(1-V)(1-y) - 2yV\ . 
\end{equation}

The relationship between the variables $\{U,V\}$ and $\{\Sigma,K\}$ 
is given by
  \begin{eqnarray}
  \label{eq:UVSK}
    U &=& \frac{(1-\Sigma^2-K)(1-y)}{(1-\Sigma^2-K)(1-y)+Ky-(1-\Sigma)^2y}
                 \label{eq:USK} \ , \mathletter{a}\\ 
    V &=& \frac{(K-(1-\Sigma)^2)(1-y)}{K-(1-\Sigma)^2+2y}
                 \label{eq:VSK} \ . \mathletter{b}
  \end{eqnarray}

\section{Lane-Emden variables}
\label{sec:LEtrf}

In the Lane-Emden approach, dimensionless
variables variables $\theta_{\rm LE},v_{\rm LE}$, and $\xi_{\rm LE}$
are introduced according to 
\begin{equation}
  \theta_{\rm LE} = \left( \frac{\rho}{\rho_c} \right)^{1/n} \ , \quad
  v_{\rm LE} = \frac{A^3m(r)}{4\pi\rho_c}\ , \ \xi_{\rm LE} = Ar\ ,
\end{equation}
where $A$ is a constant defined by
\begin{equation}
  A = \sqrt{\frac{4\pi(1-y_c)\rho_c}{(1+n)y_c}}\ ,
\end{equation}
(see, for example, Tooper \cite{art:Tooper1964}). 

The relationship to the variables $\left\{\Sigma,K,y\right\}$ is
given by
  \begin{eqnarray}
    \theta_{\rm LE} &=& \frac{y(1-y_c)}{y_c(1-y)}\ , \mathletter{a}\\
    v_{\rm LE} &=& \left[
    \frac{1-\Sigma^2-K}{8(1+n)^3}\left(\frac{1-y_c}{y_c}\right)^{3-n}
    \left(\frac{1-y}{y}\right)^{1+n}
  \right]^{1/2}W \ , \mathletter{b}\\
  \xi_{\rm LE} &=& 
    \left[\frac{1-\Sigma^2-K}{2K(1+n)}\left(\frac{y_c}{1-y_c}
    \right)^{n-1}\left(\frac{1-y}{y}\right)^{1+n}\right]^{1/2}\
    , \mathletter{c}  
  \end{eqnarray}
where
\begin{equation}
W = \frac{K-(1-\Sigma)^2}{K^{3/2}}\ .
\end{equation}

The relationship to the variables $\left\{U,V,y\right\}$ is given by
  \begin{eqnarray}
    \theta_{\rm LE} &=& \frac{y(1-y_c)}{y_c(1-y)}\ , \mathletter{a}\\
    v_{\rm LE} &=&\left[
    \frac{UV^3}{(1+n)^3(1-U)(1-V)^3}\left(\frac{y(1-y_c)}{y_c(1-y)}
    \right)^{3-n}\right]^{1/2} \ , \mathletter{b}\\
    \xi_{\rm LE} &=& \left[
    \frac{UV}{(1+n)(1-U)(1-V)}\left( \frac{y_c(1-y)}{y(1-y_c)}
    \right)^{n-1} \right]^{1/2} \ . \mathletter{c}
  \end{eqnarray}







\begin{acknowledgment}
The authors wish to thank Alan Rendall and Bernd Schmidt for comments and 
for bringing important earlier work in the field to our attention.
This research was supported by G{\aa}l\"ostiftelsen (USN), Svenska
Institutet (USN), Stiftelsen Blanceflor (USN), the University of
Waterloo (USN), and the Swedish Natural Research Council (CU).
\end{acknowledgment}

\end{article}


\begin{thebibliography}{10}

\bibitem{art:Buchdahl1959}
H.~Buchdahl, {\em Phys. Rev.} {\bf 116} (1959), 1027.

\bibitem{book:Carr1981}
J.~Carr, ``Applications of center manifold theory'', Springer Verlag, 
New York, 1981.

\bibitem{book:Chandra1939}
S.~Chandrasekhar, ``An introduction to the study of stellar structure'', 
University of Chicago Press, Chicago, 1939.

\bibitem{art:deFeliceetal1995}
F.~deFelice, Y.~Yu, and J.~Fang, {\em Class. Quant. Grav.} {\bf 12} 
(1995), 739.

\bibitem{art:Goliathetal1998a}
J.~M. Goliath, U.~S. Nilsson, and C.~Uggla, {\em Class. Quant. Grav.}
{\bf 15} (!998), 167.

\bibitem{art:Goliathetal1998b}
J.~M. Goliath, U.~S. Nilsson, and C.~Uggla, {\em Class. Quant. Grav.}
{\bf 15} (1998), 2841.

\bibitem{art:Hartle1978}
J.~B. Hartle, {\em Phys. Rep.} {\bf 46} (1978), 201.

\bibitem{art:Horedt1987}
G.~P. Horedt, {\em {A}stron. {A}strophys.} (1987) 177, 117.

\bibitem{art:Kimura1981}
H.~Kimura, {\em {P}ubl. {A}stron. {S}oc. {J}apan} {\bf 33} (!981), 273.

\bibitem{art:Makino1998}
T.~Makino, {\em Journal of Mathematics of Kyoto University}
{\bf 38} (1998), 55.

\bibitem{art:MisnerSharp1964}
C.~W. Misner and D.~H. Sharp, {\em Phys. Rev.} {\bf 136} (1964), B571.

\bibitem{book:gravitation}
C.~W. Misner, K.~S. Thorne, and J.~A. Wheeler, ``Gravitation'', 
 Freeman and Co., New York, 1973.

\bibitem{art:grstar_leq}
U.~S. Nilsson and C.~Uggla, ``General relatvistic stars: Linear 
equations of state'', preprint 2000.

\bibitem{art:RendallSchmidt1991}
A.~D. Rendall and B.~G. Schmidt, {\em Class. Quant. Grav.} {\bf 8} (1991), 
985.

\bibitem{art:Schwarzschild1916}
K.~Schwarzschild, {\em Sitzber. Deut. Akad. Wiss. Berlin, Kl. Math.-Phys. 
Tech}, page 189, 1916.

\bibitem{art:Tooper1964}
R.~F. Tooper, {\em Astrophys. J.} {\bf 140} (1964), 434.

\bibitem{book:WainwrightEllis1997}
J.~Wainwright and G.~F.~R. Ellis, ``Dynamical systems in cosmology'', 
{C}ambridge {U}niversity {P}ress, Cambridge, 1997.

\end{thebibliography}
\end{document}